# Predicting the Impact of Generative AI Using an Agent-Based Model

Joao Tiago Aparicio
LNEC, INESC-ID, Instituto Superior Técnico, Universidade de Lisboa, Lisboa, Portugal
Joao.aparicio@tecnico.ulisboa.pt

Manuela Aparicio
NOVA Information Management School (NOVA IMS), Universidade Nova de Lisboa, Portugal
manuela.aparicio@novaims.unl.pt

Sofia Aparicio
Instituto Superior Técnico, Universidade de Lisboa, Lisboa, Portugal
sofia.aparicio@tecnico.ulisboa.pt

Carlos J. Costa
Advance/ISEG (Lisbon School of Economics & Management), Universidade de Lisboa, Lisbon, Portugal
cjcosta@iseg.ulisboa.pt

*Abstract* — **Generative artificial intelligence (AI) systems have transformed various industries by autonomously generating content that mimics human creativity. However, concerns about their social and economic consequences arise with widespread adoption. This paper employs agent-based modeling (ABM) to explore these implications, predicting the impact of generative AI on societal frameworks. The ABM integrates individual, business, and governmental agents to simulate dynamics such as education, skills acquisition, AI adoption, and regulatory responses. This study enhances understanding of AI's complex interactions and provides insights for policymaking. The literature review underscores ABM's effectiveness in forecasting AI impacts, revealing AI adoption, employment, and regulation trends with potential policy implications. Future research will refine the model, assess long-term implications and ethical considerations, and deepen understanding of generative AI's societal effects.**

## I. Introduction

Generative artificial intelligence (AI) systems have gained significant traction in recent years [21], revolutionizing various industries such as healthcare, finance, and entertainment. These systems can autonomously generate content, including text, images, audio, and video, mimicking human-like creativity. While generative AI holds immense promise in enhancing [22] productivity and innovation, its widespread adoption raises concerns about its potential social and economic consequences. Therefore, there is a pressing need to understand and predict the implications of generative AI within societal frameworks.

In this paper, we aim to explore the social and economic consequences of using generative AI through the lens of agent-based modeling (ABM). Our objectives include proposing a possible agent-based model to simulate the impact of generative AI adoption and its interactions within socio-economic systems. This research started with a comprehensive literature review to identify existing research on generative AI's social and economic implications. Then, an agent-based model was proposed that captures the complex dynamics of generative AI adoption and its effects on various societal factors. It implemented the proposed model and validated its efficacy in predicting the social and economic consequences of using generative AI. Finally, the simulations' results are presented, and the potential implications for policymaking and decision-making processes are analyzed.

Our approach begins with a thorough literature review to gather insights into generative AI's social and economic impacts. Building on existing research, we propose an agent-based model that incorporates key variables and relationships relevant to adopting and utilizing generative AI within society. We then implement the model using appropriate simulation techniques and validate its performance against real-world data where possible. Finally, we present the results of our simulations, highlighting key findings and discussing their implications for stakeholders and policymakers.

Our study aims to contribute to a better understanding of the complex interactions between generative AI and socio-economic systems, providing valuable insights into the potential consequences of its widespread adoption. Through agent-based modeling, we seek to offer a predictive framework for assessing the social and economic implications of using generative AI, thereby informing decision-making processes and facilitating the development of policies that balance innovation with societal well-being.

## II. Literature Review

Agent-based modeling (ABM) emerges as a robust methodology for predicting the multifaceted impact of generative artificial Intelligence (AI) on socio-economic systems [1] [2]. This innovative approach facilitates the simulation of micro-level agents whose interactions give rise to macroscopic patterns aligned with predetermined objectives, providing insights into the intricate dynamics inherent in generative AI systems [2] [3]. By reverse-engineering desired outcomes, researchers can unravel the underlying rules and parameters governing these agents, thereby enhancing our understanding of AI's societal implications [2].

Furthermore, ABM holds promise for evaluating the social repercussions of generative AI systems across diverse modalities, including text, images, audio, and video [4] [5]. By specifying fundamental agent-rule components and permissible combinators, ABM enables a nuanced analysis of various dimensions such as biases, stereotypes, representational harms, cultural values, and disparate performance [5]. Additionally, ABM facilitates the assessment of financial costs, environmental impacts, and labor costs associated with data and content moderation, contributing to a comprehensive understanding of AI's socio-economic ramifications [5].

Leveraging insights from ABM empowers researchers to make informed decisions regarding integrating generative AI into societal frameworks [6] [7]. Comprehensive simulations and analyses, supported by empirical evidence, enable stakeholders to anticipate and address complex challenges posed by AI technologies, thereby maximizing societal benefits while mitigating potential negative consequences [6] [7].

In conclusion, ABM is a valuable tool for forecasting and analyzing generative AI systems' social and economic impacts [8]. By simulating complex interactions and dynamics, ABM facilitates a comprehensive understanding of AI's potential societal consequences, informing strategic decision-making and policy development [9].

## III. PROPOSING A SET OF AGENTS

In our agent-based model, we aim to predict the impact of Generative AI adoption on various societal factors such as education, skills, employment, and government regulation. The model incorporates three primary types of agents: individual agents, business agents, and government agents. The dynamics of each agent are described by differential equations representing their continuous-time evolution. This model is based on design science research [19] and follows a similar architecture to previous multi-agent systems studies [20].

### A. Individual Agents:

The dynamics for individual agents involving education and skill acquisition follow logistic growth models and differential equations commonly used in educational and psychological studies [10]. The sigmoid function used for modeling skill acquisition represents the concept of diminishing returns, a principle well-documented in educational theory and cognitive psychology [11,12]. Each agent seeks education ($E(t)$) and acquires skills ($S(t)$). Education is influenced by the rate of seeking ($\alpha$), and skills are influenced by education and the learning rate ($\beta$). The equations governing individual agents are:

$dE/dt = \alpha \cdot (1 - E(t))$

$dS dt = \beta \cdot E(t) S(t) = 1 1 + e - \beta \cdot E(t),$

*then normalized between 0 and 1*

Applying a sigmoid function to model skill acquisition over time introduces a non-linear growth pattern that reflects realistic learning curves[23]. Early on, as education increases, skills proliferate rapidly. However, as skills approach the maximum limit (asymptote of the sigmoid function), additional education results in smaller increases in skill [24]. This approach models the psychological principle of diminishing returns in learning, where initial education leads to significant skill gains, but these gains taper off as one becomes more skilled [23].

The sigmoid function can simulate scenarios where individuals continue to learn and develop skills, but the impact of further education is less pronounced as they become more skilled[4]. This is a more realistic representation compared to linear models, as it captures the inherent non-linearity in skill acquisition over time[23].

The sigmoid function has been widely used in various fields to model growth patterns that exhibit an S-shaped curve[4]. In plant growth studies, the beta growth function, a flexible sigmoid function, has been used to successfully describe the sigmoid dynamics of seed filling, plant growth, and crop biomass production[24]. The beta growth function has clearly interpretable parameters and can specify the length of the growth period and smoothly predict the final weight of the determinate growth process[24].

In summary, applying a sigmoid function to model skill acquisition over time introduces a realistic non-linear growth pattern that reflects the psychological principle of diminishing returns in learning. This approach can simulate scenarios where individuals continue to learn and develop skills, with the impact of further education diminishing as they become more skilled..

### B. Business Agents:

Business agents' behavior, particularly their adoption of generative AI based on available skills, can be modeled by logistic growth influenced by internal learning rates, echoing theories from business management and economics regarding technology adoption [13]. Businesses adopt Generative AI ($A(t)$) based on the skills possessed by individual agents. The adoption rate is influenced by the learning rate of businesses ($\gamma$).

The equation governing business agents is:

$dA/dt = \gamma \cdot (1 - A(t)) \cdot S(t)$

### C. Labor Market (Employment):

Integrating skills' probability density functions into labor market dynamics uses statistical methods typical in labor economics, especially for modeling labor supply and demand shifts due to technological change [14]. Let $f(S, t)$ represent the skills' probability density function (PDF) at time t, where S is the skill level. The total supply of labor (Total Supply Labor(t)) is given by the integral of the product of skill level and probability density function:

Total Supply Labor(t) = $\int_{[-\infty,\infty]} S \cdot f(S, t) \cdot dS$

The total demand factor (Total Demand Factor(t)) is influenced by the adoption of Generative AI ($A(t)$) and other potential factors:

Total Demand Factor(t) = $\gamma \cdot (1 - A(t))$

The dynamics of employment are governed by the rate of change of the integral representing the minimum of total supply labor, total demand factor, and the maximum supply labor threshold (Max Supply Labor):

dEmployment/dt = $d/dt[\min(\int_{[-\infty,\infty]} S \cdot f(S, t) \cdot dS, \gamma \cdot (1 - A(t)),$ Max Supply Labor)]

### D. Government Agent:

The modeling of government regulation as a response to AI adoption levels can draw on theories from regulatory economics and public policy, particularly those discussing the feedback mechanisms between industry behavior and regulatory adjustments [15]. The government regulates AI development ($R(t)$) based on the level of Generative AI adoption. The

regulation rate is influenced by a regulation factor (δ). The equation governing the government agent is:

$$dR/dt = \delta \cdot (A(t) - R(t))$$

*E. Model Dynamics and Predictions:*

Concerning education and skills, individual agents continuously seek education, leading to an increase in skills over time. This reflects the ongoing learning and skill development in the population.

Concerning generative AI adoption, Businesses adopt Generative AI at a rate influenced by the population's skills. As skills increase, businesses' adoption of AI is expected to grow.

Employment is a dynamic variable influenced by both the supply and demand for labor. The model predicts how the adoption of Generative AI impacts the balance between labor supply and demand, affecting overall employment levels.

The government regulates AI development based on the level of AI adoption. This regulation is a feedback mechanism that aims to control and manage the societal impact of AI.

Differential equations and integrals allow for a continuous-time representation of agent behaviors and their interactions, providing insights into the long-term effects of Generative AI adoption on various societal aspects.

## IV. IMPLEMENTATION

We have implemented the proposed agent-based model using Python, leveraging libraries such as NumPy and Matplotlib to facilitate its development. The model architecture revolves around four main classes: IndividualAgent, BusinessAgent, LaborMarket, and GovernmentAgent.

IndividualAgent is a class representing agents engaged in seeking education and supplying labor. BusinessAgent reflects businesses adopting generative AI technology and influencing economic labor demand. LaborMarket is a class that Facilitates the matching of labor supply and demand, serving as the dynamic nexus of the labor ecosystem. GovernmentAgent is a class that governs the regulation of AI development based on prevailing AI adoption levels, shaping the technological landscape.

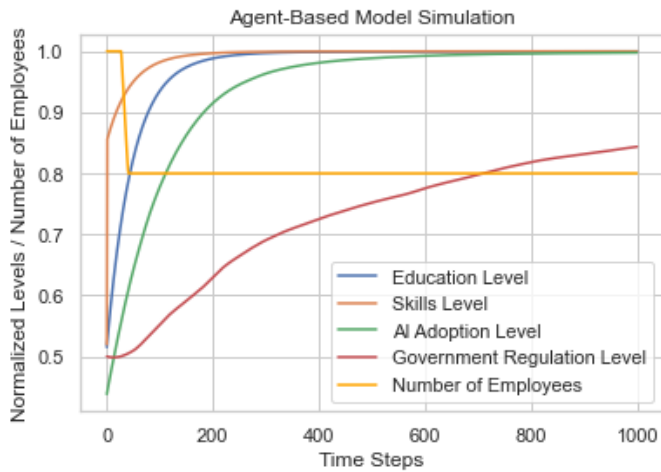

Figure 1. Small Agent-Based Model Simulation.

Upon initialization, the model is equipped with learning rates, adoption rates, and regulatory factors. We establish initial conditions for education, skills, AI adoption, and regulation before commencing simulation over multiple time steps, enabling observation of its behavior and prediction of outcomes. Figure 1 shows a small-agent community simulation.

The simulation outcomes provide invaluable insights into generative AI adoption's potential social and economic ramifications. Notable trends include businesses' escalating adoption of AI as individuals acquire skills, thereby influencing employment levels and government regulation. The model is a versatile tool for exploring diverse scenarios and conducting thorough analyses of parameter impacts on outcomes.

From a software engineering perspective, our code adheres to best practices in design and implementation. Embracing an object-oriented paradigm, we organize functionality into cohesive classes—IndividualAgent, BusinessAgent, LaborMarket, GovernmentAgent, and AgentBasedModel—enhancing readability and modularity. This architectural approach promotes scalability and maintainability by encapsulating related behavior and attributes within well-defined abstractions.

Each class's constructor methods (init) are central in initializing object attributes and fostering encapsulation and abstraction. This design decision facilitates code reusability and clarity by encapsulating object state and behavior within self-contained units. The simulate method within the AgentBasedModel class encapsulates the simulation logic, ensuring separation of concerns and modularity. This modular structure allows for seamless modification, extension, or replacement of simulation components without disrupting other model facets.

Additionally, the set_initial_conditions method empowers users to specify initial parameters and conditions, augmenting simulation flexibility and configurability. By decoupling initialization logic from the primary simulation loop, this method facilitates experimentation with diverse scenarios and parameter values, bolstering sensitivity analysis and model validation efforts.

Random variability in alpha, beta, and gamma parameters imbues the simulation with stochasticity, mirroring real-world uncertainties and heterogeneities. This stochastic element enriches the model's realism, enabling the exploration of probabilistic outcomes and scenarios.

Furthermore, error handling mechanisms are implemented to safeguard against negative values or invalid states in specific attributes, such as demand_factor and regulation. These defensive programming practices fortify simulation stability and correctness, ensuring robustness and reliability in results.

In summary, our agent-based model epitomizes rigorous software engineering principles and design patterns, harnessing the versatility of Python and the robust capabilities of libraries like NumPy and Matplotlib. It offers a flexible, robust, and interpretable framework for in-depth analysis of labor market dynamics, facilitating comprehensive studies at the intersection of economics, education, and technology adoption.

## V. RESULTS

The model presented may give some insights into education, skills, AI adoption, government regulation, and employment levels. This is shown in Figure 2 (next page), where we use a large community (100k) of agents over 100 simulations and 1k timesteps.

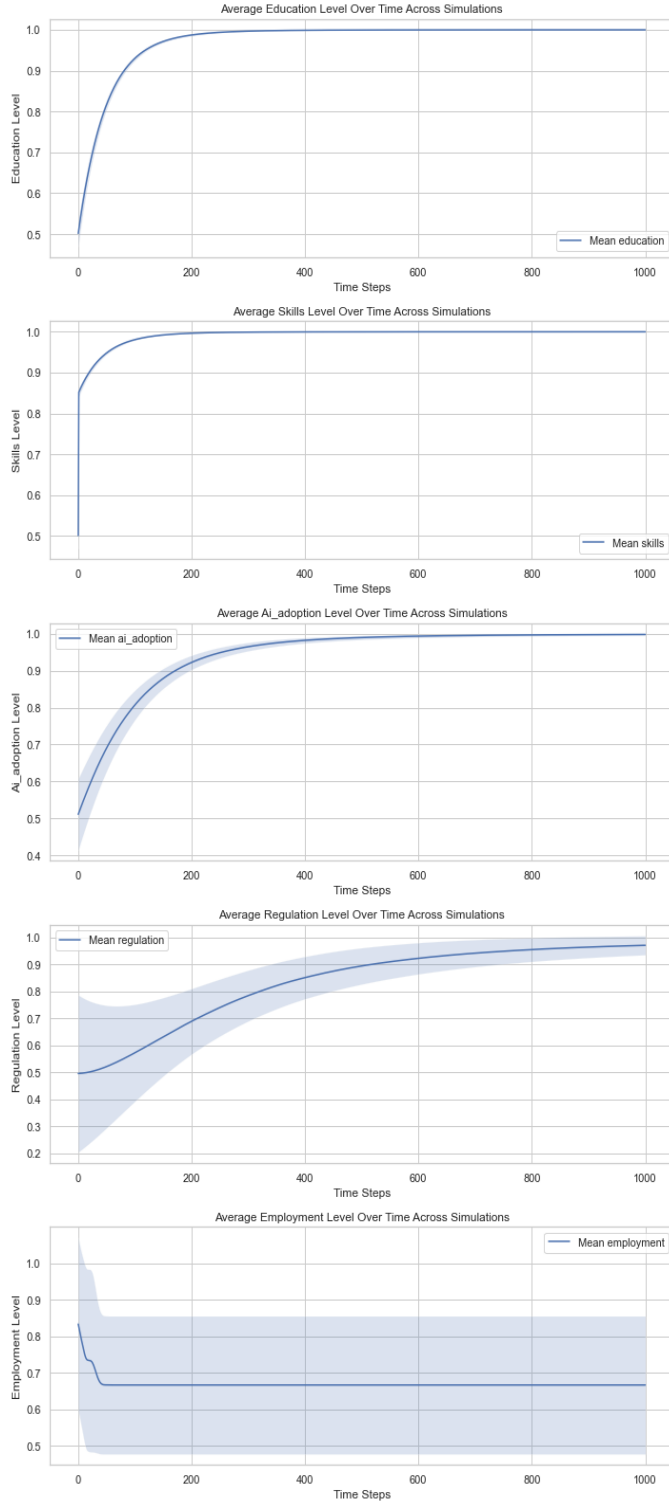

Figure 2. Agent-Based Model Simulation.

The agent-based model presented in the search results provides several insights into the dynamics of education, skills, AI adoption, government regulation, and employment levels: Education Level, Skills Level, AI Adoption, Regulation Level, and Employment Level.

The model shows a logistic growth pattern in education levels, quickly reaching a saturation point. This suggests diminishing returns to investment in education or other factors limiting further growth once a certain education level is reached, consistent with current literature [16, 25,26].

The skills level derived from education follows a logistic growth curve, indicating that as individuals become more educated, their skills increase at a diminishing rate, approaching an upper limit. This could reflect real-life scenarios where further education results in smaller incremental increases in practical skills [27,28].

AI adoption grows over time, suggesting that businesses continuously adopt more AI technology as they respond to the skills available in the labor market, in line with existing research [18, 29,30].

The regulation level appears to be increasing steadily, potentially as a government's response to rising AI adoption, aiming to manage the impact on the labor market and society. The model suggests a lag or gradual approach to regulation, which may be due to bureaucratic inertia or a deliberate phased implementation [31,32].

The employment level initially decreases, possibly due to AI adoption's impact on labor demand, but then stabilizes and does not drop below a certain threshold, potentially due to social safety nets or policies that maintain a minimum level of employment regardless of technological impact [17, 33,34].

## VI. DISCUSSION

The agent-based model offers a valuable framework for exploring the complex dynamics between education, skills, AI adoption, regulation, and employment. While the results align with existing literature, the model provides additional insights and implications for policymakers.

Regarding education and skills, the observed logistic growth patterns confirm studies showing diminishing returns on education investment and the need to focus on developing practical, in-demand skills [25,26,27,28]. This suggests that policymakers should prioritize education reforms to better align curricula with the skills required in an AI-driven economy, such as integrating data science and machine learning courses.

The continuous adoption of AI technology by businesses, driven by available skills [29,30], highlights the importance of supporting ongoing skill development and lifelong learning initiatives. Policies encouraging retraining and upskilling programs can help workers adapt to technological changes and maintain employability.

The model's insights on the gradual and lagging approach to regulation [31,32] emphasize policymakers' challenges in keeping pace with rapid technological advancements. Developing flexible regulatory frameworks that balance innovation with risk mitigation will be crucial in managing the societal impacts of AI.

Finally, the model's findings on employment levels [33,34] stress the need for robust social safety nets and policies to mitigate potential disruptions caused by AI-driven automation. Proactive measures, such as retraining programs and income support, can ease the transition for workers displaced by technological change.

Overall, the agent-based model presented provides a valuable tool for exploring the interplay between education, skills, AI adoption, regulation, and employment. By comparing the model's results with existing academic literature, policymakers can comprehensively understand these dynamics and develop informed strategies to navigate the challenges and opportunities presented by the increasing integration of AI in the economy and society.

In practical terms, these trends inform policymakers about several crucial aspects. First, ensuring that education systems prioritize acquiring skills demanded in an AI-driven economy is imperative. Integrating data science and machine learning courses into curricula can better prepare students for future job markets. Second, monitoring AI's impact on employment is essential, with policies ready to support those affected by technological change. Establishing retraining programs for workers displaced by automation can facilitate their transition into new roles or industries. Third, planning for a future with prevalent AI requires adaptable regulations that keep pace with technological advancements. Creating flexible frameworks that encourage innovation while safeguarding against potential risks associated with AI deployment is necessary. Lastly, supporting continuous learning and skill development is critical for ensuring ongoing employability in a rapidly evolving job market. Initiatives such as lifelong learning grants or subsidies for upskilling programs can help individuals stay competitive amidst technological shifts. These actions address immediate challenges and prepare societies for a future where AI significantly shapes economies and industries.

## VII. CONCLUSIONS

This paper describes an agent-based model (ABM) to predict the social and economic consequences of using generative artificial Intelligence (AI). Through simulations conducted using the proposed model, we have explored the potential impacts of generative AI adoption on various societal factors, including education, employment, and government regulation. Our simulations have provided insights into the dynamics of generative AI adoption and its interactions within socio-economic systems. We have observed trends such as businesses' increasing adoption of AI as the population's skills improve and the potential implications for employment levels and government regulation. Key Findings may be considered at the following level: Education and Skills Development, AI Adoption Trends, Regulatory Dynamics, and Employment Impact. The simulation revealed a logistic growth in education and skills levels, pointing towards a saturation point beyond which additional education does not proportionally enhance skills. This suggests optimizing educational resources and strategies to maximize effective learning and skill acquisition without unnecessary overextension. AI adoption increased steadily, influenced by the availability of skilled labor and regulatory frameworks. This underscores the importance of aligning educational outcomes with market needs, ensuring that the labor force is equipped to handle and benefit from emerging AI technologies. The gradual increase in regulatory measures in response to AI adoption highlights the necessity for governments to adopt a proactive and phased approach to regulation. This strategy accommodates technological advancements while ensuring societal impacts are managed and mitigated. The initial decline followed by stabilization in employment levels indicates the disruptive impact of AI on traditional jobs, countered by employment policies possibly designed to uphold a minimum level of workforce engagement. This aspect of the findings stresses the importance of robust social safety nets and policies adapting to technological disruptions.

Based on the findings of this study, we recommend the following actions for stakeholders: Educational Policy Reform, Strategic AI Integration, Proactive Regulatory Frameworks, Support for Displaced Workers, and Lifelong Learning Systems. Policymakers should focus on revising educational curricula to emphasize skills directly relevant to an AI-driven economy. This involves technical skills related to AI and data science and adaptive skills such as problem-solving, critical thinking, and lifelong learning capabilities. Businesses should strategically adopt AI technologies, focusing on augmenting rather than replacing human labor where possible. Investment in employee training to work alongside AI will maximize productivity and innovation. Governments should develop flexible, forward-looking regulatory frameworks that can quickly adapt to new developments in AI technology. This involves regular dialogue with technologists, business leaders, and academics to anticipate future trends and potential impacts. There is a pressing need for policies that support workers displaced by AI, such as retraining programs, unemployment benefits, and career counseling services. These programs should be designed to quickly and effectively re-integrate workers into the evolving job market. The development of continuous education and training systems is essential. These systems should be accessible to all workers throughout their careers, allowing them to continuously update their skills and stay relevant in a rapidly changing economic landscape. However, it is essential to note that our model relies on certain assumptions and parameter values that may not fully capture the complexity of real-world scenarios. In future work, there is a need to refine and validate the model by estimating parameters based on real-life statistics and data. We can enhance the model's predictive accuracy and reliability by calibrating the model to fit empirical observations better.

Additionally, future research could explore the long-term implications of generative AI adoption, considering technological advancements, societal attitudes, and regulatory changes. Furthermore, there is a need to assess the potential ethical and moral implications of using generative AI, including bias, privacy, and accountability issues. Our study represents an initial step toward understanding and predicting generative AI's social and economic consequences. Refining and expanding our model in future research can further advance our understanding of this rapidly evolving technology and its impact on society.

ACKNOWLEDGMENT

We gratefully acknowledge financial support from FCT - Fundação para a Ciência e a Tecnologia (Portugal), national funding through research grant UIDB/04521/2020. National funds also support this work through PhD grant (UI/BD/153587/2022) supported by FCT.